\documentclass{article}%
\usepackage{amsmath}%
\usepackage{amsfonts}%
\usepackage{amssymb}%
\usepackage{graphicx}
\begin{document}

\title{DOES CHARGE CONTRIBUTE TO THE FRAME DRAGGING OF SPACETIME?}
\author{Babur M. Mirza\\Department of Mathematics, \\Quaid-i-Azam University, Islamabad. \\45320. Pakistan\\(E-mail:bmmirza2002@yahoo.com)}
\maketitle

\begin{abstract}
Electrically charged systems bound by a strong gravitational force can sustain
a huge amount of electric charge (up to $10^{20}C$) against Coulomb repulsion.
General relativistically such systems form a stable hydrostatic configuration
both in the non-rotating and rotating cases$.$ Here we study the effects of
electric charge (electric energy density) on the spacetime outside a rotating
electrically charged system bound by a strong gravitational force. In
particular we investigate the effect of charge density on frame-dragging of
spacetime in the exterior region. Using the coupled Einstein-Maxwell equations
it is found that in the slow rotation approximation charge accumulations not
only acts like an additional mass, thus modifying the spherically symmetric
part of the spacetime, the electric charge also contributes directly to the
dragging of spacetime. A modified Lense-Thirring formula for the spacetime
frame dragging frequency is obtained.

\end{abstract}

\section{INTRODUCTION}

Hydrostatic equilibrium in physical systems bound by Newtonian gravitation is
achieved essentially under electrically neutral conditions$^{1}$. A comparison
between the repulsive Coulomb force and the attractive gravitational force
shows that for systems such as the Sun, charge accumulation cannot be
significantly larger than $10^{2}C$. For systems with intense gravitational
field such as compact stars the Newtonian theory does not lead to correct
equilibrium conditions, here modifications in the gravitational field
equations consistent with the general theory of relativity must be introduced.
This leads to the general relativistic hydrostatic conditions (the
Tolman-Oppenheimer-Volkoff or TOV equations) for equilibrium in relativistic
stars. The relativistic hydrostatic equilibrium conditions for electrically
charged systems imply that charge contained by these systems can be very
large. It is estimated that both in non-rotating as well as rotating
cases$^{2-8}$ the amount of net charge contained in a compact charged star can
be as high as $10^{20}C$. \ In case of a non-rotating charged compact star
(electar) the spacetime is modified by the electrical energy density. Here the
effect of charge appears as an addition to the total mass of the star(system),
hence the its energy density. For rotating gravitational systems
frame-dragging of spacetime is a typically relativistic effect with no
analogue in classical theories of gravitation$^{9}$. In the case of a neutral
massive object, a neutron star for instance, the spacetime dragging effects
both material as well as radiative processes in vicinity of the star.

In this paper we study the effect of charge on the spacetime outside a
rotating electrically charged relativistic system, such as a charged compact
star. In particular we ask if the dragging of spacetime influenced by the
presence of electric charge on a relativistic star. In section 2 the problem
is formulated assuming a slowly rotating mass with an axially symmetric metric
describing the spacetime around the gravitational source. The form of the
metric is then determined using Einstein field equations coupled to the
Maxwell equations. This leads to a modified formula for Lense-Thirring
spacetime dragging frequency, given in section 3. In section 4 we discuss the
main conclusions and give a summary of the paper.

\section{GENERAL RELATIVISTIC FORMULATION}

We consider an electrically charged, slowly rotating star of total mass $M$
and radius $R$ enclosing a net electric charge $Q$. The spacetime exterior to
the star is described by the metric%

\begin{equation}
ds^{2}=e^{\nu(r)}dt^{2}-e^{\lambda(r)}dr^{2}-2\omega(r)r^{2}\sin^{2}\theta
dtd\varphi-r^{2}(d\theta^{2}+\sin^{2}\theta d\varphi^{2}),
\end{equation}
where $\omega(r)$ is the dragging frequency of a local inertial frame at the
location $r$ outside the star. The spacetime exterior to the charged star will
be explicitly determined using the coupled Einstein-Maxwell equations,%

\begin{equation}
R_{\beta}^{\alpha}-\frac{1}{2}\delta_{\beta}^{\alpha}R=-8\pi T_{\beta}%
^{\alpha},
\end{equation}
supplemented with the stress energy-momentum tensor for the electromagnetic
field given by%

\begin{equation}
T_{\beta}^{\alpha}=\frac{1}{4\pi}(F^{\alpha\gamma}F_{\gamma\beta}-\frac{1}%
{4}\delta_{\beta}^{\alpha}F_{\mu\nu}F^{\mu\nu}).
\end{equation}
where the electromagnetic field tensor $F^{\mu\nu}$ is given by the Maxwell equations:%

\begin{equation}
(\sqrt{-g}F^{\alpha\beta}),_{\beta}=4\pi\sqrt{-g}J^{\alpha}.
\end{equation}
For electric charge density $\rho_{e}(r)$ we define the electric field $E(r) $ as%

\begin{equation}
E(r)=\frac{4\pi}{r^{2}}\int_{0}^{r}\rho_{e}(r^{\prime})r^{\prime2}%
e^{\lambda(r^{\prime})/2}dr^{\prime},
\end{equation}
then for the components $T_{t}^{t}$ and $T_{r}^{r}$ the stress tensor can be
expressed as%

\begin{equation}
\frac{1}{4\pi}(F^{\alpha\gamma}F_{\gamma\beta}-\frac{1}{4}\delta_{\beta
}^{\alpha}F_{\mu\nu}F^{\mu\nu})=-\frac{E^{2}}{8\pi};\quad\alpha=\beta=0,1,
\end{equation}
and $E^{2}/8\pi=E(R)^{2}/8\pi$ is the electric energy density of the star. For
$r>R$ the spherical symmetric parts of the Einstein field equation coupled to
the Maxwell equation give%

\begin{equation}
\frac{e^{-\lambda}}{r^{2}}(r\frac{d\lambda}{dr}-1)+\frac{1}{r^{2}}=E^{2},
\end{equation}
\begin{equation}
\frac{e^{-\lambda}}{r^{2}}(r\frac{d\nu}{dr}+1)-\frac{1}{r^{2}}=-E^{2},
\end{equation}
The first of Einstein field equations gives the metric component
$e^{-\lambda(r)}$ as%

\begin{equation}
e^{-\lambda(r)}=1-\frac{8\pi}{r}\int_{0}^{r}\rho r^{2}dr=1-\frac{2M(r)}%
{r},\quad r>R
\end{equation}
where
\begin{equation}
M(r)=4\pi\int_{0}^{r}(\rho(r)+\frac{E^{2}}{8\pi})r^{2}dr,
\end{equation}
is the modified mass. Also adding these two equations together and solving the
resulting expression we get $\nu(r)=-\lambda(r)$. Corresponding to the axially
symmetric part of the Einstein field equations $G_{\varphi}^{t}=8\pi
T_{\varphi}^{t}$ we obtain the following equation
\begin{equation}
e^{-\lambda(r)}(r\frac{d^{2}\omega}{dr^{2}}+4\frac{d\omega}{dr})=16\pi
rE^{2}\omega.
\end{equation}
In terms of the net charge $Q=4\pi\int_{0}^{R}\rho_{e}r^{2}e^{\lambda(r)/2}dr$
the last equation can be expressed as
\begin{equation}
\frac{d}{dr}(r^{4}\frac{d\omega(r)}{dr})=16\pi Q^{2}\omega(r)e^{\lambda(r)}%
\end{equation}
where $e^{\lambda(r)}=$ $r/(r-2M)$. In this equation the dragging frequency
outside the star is modified by the electrical energy due the presence of a
net electric charge in the star.

\section{MODIFIED LENSE-THIRRING FRAME DRAGGING FREQUENCY}

Equation (12) cannot be solved exactly, however for $M/r\ll1$ it is possible
to obtain a physically interesting solution to (12), such as for the case of a
charged compact star. Under this approximation equation (12) becomes\ \ \ \
\begin{equation}
\frac{d^{2}\omega}{dr^{2}}+\frac{4}{r}\frac{d\omega}{dr}-\frac{q}{r^{4}}%
\omega=0,
\end{equation}
where we define $q\equiv16\pi Q^{2}$. We notice that equation (13) has a
regular singular point at $r$ whereas equation (12) has a removable
singularity at $r=2M$ also. Solution to equation (13) can be obtained by
standard integration procedure, which gives%

\begin{equation}
\omega(r)=\frac{r-q}{r}C_{1}e^{q/r}-\frac{r+q}{2q^{3}r}C_{2}e^{-q/r},\quad
q\neq0,
\end{equation}
where $C_{1}$ and $C_{2\text{ }}$are constants of integration. Equation (14)
is only an approximate solution of the original equation (12). However a
comparison between numerical solutions to equation (12) and solution (14)
shows that solution matches correctly to the exact solution for $M/r<1$. A
drawback of the solution (14) seems to be its non-reducibility to the case of
an uncharged electrically neutral object, that is the limit when
$q\rightarrow0$. To investigate this more fully we expand the exponentials in
(14) in a power series, after some simplification we have
\begin{equation}
\omega(r)=C_{1}(1-\frac{q^{2}}{2r^{2}}-\frac{q^{3}}{3r^{3}}-...)-C_{2}\frac
{1}{2q^{3}}(1-\frac{q^{2}}{2r^{2}}+\frac{q^{3}}{3r^{3}}-+...),\quad q\neq0,
\end{equation}
separating terms independent of $q$ we obtain%

\begin{equation}
\omega(r)=-\frac{C_{2}}{3r^{3}}+C_{1}-\{C_{1}(\frac{q^{2}}{2r^{2}}+\frac
{q^{3}}{3r^{3}}+...)+C_{2}(\frac{1}{2q^{3}}-\frac{1}{4qr^{2}}+-...)\},\quad
q\neq0.
\end{equation}
Here the first two term gives the correct solution to equation (14) for the
chargeless case. For the remaining terms in equation (16) it can be seen that
as $q$ approaches zero, the terms in parentheses\ together become vanishingly
small for a fixed value of $r$.

In the figure (1) we give exact numerical solutions of equation (12) for
different values of $q$. For comparison with the approximate formula (14) for
modified Lense-Thirring frequency, plots are given in figure (2), between the
dragging frequency $\omega(r)$ and the radial distance $r$. In both cases we
take $\omega(1)=1=\omega^{\prime}(r)$ and $M=0.2$ in gravitational units
$G=1=c$. Figure (3) consists of plots for the spherically symmetric part of
the metric $e^{-\lambda(r)}$ as a function of distance $r$ for various values
of net charge $q$ inside the star. We notice that the modified Lense-Thirring
frequency has the same profile in each case for different values of charge $q$.

\section{SUMMARY AND CONCLUSIONS}

We have studied the spacetime around a rotating electrically charged massive
object within the context of the general theory of relativity. The spacetime
metric is determined using the coupled set of Einstein-Maxwell field equations
for the case of a rotating electrically charged system. These equations are
then solved to obtain analytical expressions for the metric components.

The solutions obtained show that the presence of charge effects the spacetime
in a twofold way. Firstly, it modifies the total energy density hence the
total mass of the system. This in turn effects the spherically symmetric part
of the spacetime metric (figure 3). Secondly, the axially symmetric part of
the metric due to the rotation of the system modifies the Lense-Thirring
spacetime dragging around the gravitating object(figure 1 and 2). We obtain an
analytical formula for the modified Lense-Thirring dragging frequency which is
applicable to the case of charged compact objects under the approximation
$M/r\ll1$. We find that charged massive object (such as a charged compact
star) is not only more strongly gravitating due to the enhanced total mass
(energy density), but if rotating the dragging of spacetime around the massive
object is modified also.

This effect has important implications for electrically charged
gravitationally bound stellar systems. For instance a free falling inertial
observer (ZAMO) with velocity four vector $u_{ZAMO}=(e^{\nu(r)},0,0,\omega
(r)e^{\nu(r)})$ circling the star at fixed radial distance will not only need
larger potential energy to keep itself in orbit, but also will revolve faster
due to the enhanced dragging frequency. Estimated charge contained in an
electrically compact star can be as high as $10^{20}C$. This corresponds to
the mass increase of about ten to twenty percent mass of the star in a neutral
condition$^{6-9}$. Observationally the detection of such effects, for instance
around a charged electric star (electar), seems therefore quite feasible.

FIGURE\ CAPTIONS:

Figure 1: Plots between the spacetime dragging frequency $\omega(r)$ and the
radial distance $r$ using equation (12) for different values of the net charge
$q$.

Figure 2: Plots between the spacetime dragging frequency $\omega(r)$ and the
radial distance $r$ using the modified Lense-Thirring formula (14) for
different values of the net charge $q$.

Figure 3: Plots between the modified metric component $e^{-\lambda(r)}$ and
the radial distance $r$ using equation (9) for the exterior region with
different values of the net charge $q$.


\begin{thebibliography}{9}                                                                                                %
\bibitem {[1]}N.K. Glendenning, \textit{Compact Stars: Nuclear Physics,
Particle Physics, and General Relativity.} (Springer-Verlag; 2000); also F.
Weber, and N.K. Glendenning, \textit{Phys. Lett.} \textbf{B265}, 1 (1991).

\bibitem {[2]}J.L. Zhang, W.Y. Chau, and T.Y. Deng, \textit{Astrophys. and
Space Sc}. \textbf{88}, 81 (1982).

\bibitem {[3]}F. de Felice, Y.Q. Yu, , and L.Z. Fang, \textit{MNRAS}.
\textbf{277}, L17 (1995).

\bibitem {[4]}F. de Felice, S.M. Liu, and Y.Q. Yu, \textit{Class. and Quantum
Grav}. \textbf{16}, 2669 (1999).

\bibitem {[5]}Y.Q. Yue, and S.M. Liu, \textit{Comm. Theor. Phys.} \textbf{33},
571 (2000).

\bibitem {[6]}P. Anninos, and T. Rothman, \textit{Phys. Rev.} \textbf{D65},
024003 (2001).

\bibitem {[7]}S. Ray, A.L. Espindola, M. Malheiro, J.P.S. Lemos, and V.T.
Zanchin, \textit{Phys. Rev. }\textbf{D68}, 084004 (2003).

\bibitem {[8]}S. Ray, M. Malheiro, J.P.S. Lemos, V.T. Zanchin, \textit{Braz.
J. Phys.} \textbf{34, }310 (2004).

\bibitem {[9]}N. Stergioulas, \textit{Living Rev. Relativity}, \textbf{6}, 3 (2003).
\end{thebibliography}
\end{document}